\documentclass[twocolumn,nofootinbib,preprintnumbers,amsmath,superscriptaddress]{revtex4}

\usepackage{epsfig}
\usepackage{amssymb}
\usepackage{color}

\newcommand\eqn[1]{(\ref{#1})}      

\newcommand{\beq}{\begin{equation}}
\newcommand{\eeq}{\end{equation}}
\newcommand{\ba}{\begin{array}}
\newcommand{\bea}{\begin{eqnarray}}
\newcommand{\ea}{\end{array}}
\newcommand{\eea}{\end{eqnarray}}

\begin{document}


\title{Scenario for quark confinement from infrared safe Yang-Mills dynamics}

\author{Marcela Pel\'aez}%
\affiliation{%
Instituto de F\'{\i}sica, Facultad de Ingenier\'{\i}a, Universidad de
la Rep\'ublica, J. H. y Reissig 565, 11000 Montevideo, Uruguay.\vspace{.1cm}}%
\author{Urko Reinosa}%
\affiliation{%
Centre de Physique Th\'eorique, CNRS, Ecole Polytechnique, \\IP Paris, F-91128 Palaiseau, France.
\vspace{.1cm}}%
\author{Julien Serreau}%
\affiliation{Universit\'e Paris Cit\'e, CNRS, AstroParticule et Cosmologie, \\F-75013 Paris, France.\vspace{.1cm}}
\author{Matthieu Tissier}\affiliation{Sorbonne
Universit\'e, CNRS, Laboratoire de Physique Th\'eorique de la Mati\`ere Condens\'ee, \\75005 Paris, France.
\vspace{.1cm}}%
\author{Nicol\'as Wschebor}%
\affiliation{%
 Instituto de F\'{\i}sica, Facultad de Ingenier\'{\i}a, Universidad de
 la Rep\'ublica, J. H. y Reissig 565, 11000 Montevideo, Uruguay.\vspace{.1cm}}%

\date{\today}

\begin{abstract}
 We revisit the non-Abelian dipole problem in the context of a simple semiclassical approach that incorporates some essential features of the infrared sector of Yang-Mills theories in the Landau gauge, in particular, the fact that both the running coupling and the gluon propagator remain finite at infrared scales and that the latter shows positivity violations that reflects the presence of massless modes. We obtain a simple flux tube solution in a controlled approximation scheme, which we compare to the results of lattice simulations. 
\end{abstract}

\maketitle

\section{Introduction}

Despite tremendous progress in unraveling the various aspects of quantum chromodynamics (QCD), a definite understanding of the basic mechanism underlying the phenomenon of confinement is still lacking. In pure gauge, Yang-Mills (YM) theories, the confinement of static color sources has long been established in numerical simulations, notably through the area law for the Wilson loop or, equivalently, the linear interaction potential at large distances \cite{Wilson:1974sk,Creutz:1980zw,Bali:1992ab,Luscher:2001up}. Various scenarios have been identified for the typical field configurations responsible for the latter \cite{Greensite:2011zz}, among which the formation of a chromoelectric flux tube between the sources, which has been observed in lattice simulations~\cite{Bali:1994de,Haymaker:1994fm,Cea:1995zt}. 

Effective semiclassical descriptions aiming at modeling the specific nonlinearities of the non-Abelian dipole problem have been proposed, notably, the popular dual superconductor model \cite{Nielsen:1973cs,Mandelstam:1974pi,tHooft:1974kcl,Kogut:1974sn,Polyakov:1975rs,Ripka:2003vv}. These typically appeal to effective scalar degrees of freedom and depart from a first principle calculation. Another line of development aims at obtaining flux tube solutions from the basic equations of motion of the color fields, including the effect of quantum fluctuations~\cite{Savvidy:1977as,Adler:1981as,Adler:1982rk}. Those works,  however, rely on standard perturbation theory in a regime of long distances, where the latter is unreliable because of the diverging running coupling at infrared scales, the so-called Landau pole. 

In the present work, we revisit the question of a semiclassical description of the flux tube formation with a fresh look, taking advantage of the fact, solidly established in the last two decades, that the Landau pole is an artifact of the standard perturbative description based on the Faddeev-Popov (FP) approach \cite{Deur:2023dzc,Boucaud:2011ug}. Extensive lattice simulations of YM theories in the Landau gauge have shown that the coupling ({\it e.g.}, defined in the Taylor scheme) never diverges and even remains of moderate size \cite{Boucaud:2011ug}. This comes with another major observation, namely, the fact that the gluon field develops a nonzero screening mass of a few hundreds MeV together with positivity violations in the infrared \cite{Cucchieri:2007rg,Bogolubsky:2009dc}. Various approaches based on continuum field theory have been proposed to incorporate the relevant dynamics of the screening of the Landau pole \cite{Alkofer:2000wg,Fischer:2008uz,Dudal:2008sp,Vandersickel:2012tz,Aguilar:2013vaa,Tissier:2010ts} which have led to successful descriptions of a variety of nontrivial infrared phenomena, such as the confinement-deconfinement transition at nonzero temperature and baryonic chemical potential or the dynamical breaking of chiral symmetry in QCD \cite{Braun:2014ata,Fischer:2018sdj,Fu:2019hdw,Pelaez:2021tpq}.
We show how these {\it infrared safe} approaches allow for a controlled description of the non-Abelian dipole problem in YM theories with, in particular, simple semiclassical flux tube solutions.

\section{Setup}
In continuum approaches, it often proves convenient to work with background field techniques~\cite{Abbott:1980hw}. Specifically, denoting $A_\mu^a$ the gluon field, we introduce a background field $\bar A_\mu^a$ and we work in the Landau-DeWitt gauge defined as $(\bar D_\mu a_\mu)^a=\partial_\mu a^a_\mu+gf^{abc}\bar A_\mu^ba_\mu^c=0$, where $a_\mu^a=A_\mu^a-\bar A_\mu^a$. Here, $g$ denotes the coupling constant and $f^{abc}$ the SU($N$) structure constants. The corresponding FP gauge-fixed Euclidean Lagrangian reads
\begin{equation}\label{eq:lagFP}
 {\cal L}_{\rm FP}=\frac{1}{4}F^a_{\mu\nu}F^a_{\mu\nu}+ih^a(\bar D_\mu a_\mu)^a+(\bar D_\mu\bar c)^a(D_\mu c)^a,
\end{equation}
with $F_{\mu\nu}^a=\partial_\mu A_\nu^a-\partial_\nu A_\mu^a+gf^{abc}A_\mu^bA_\nu^c$ the field strength tensor, $D_\mu$ the covariant derivative, and $h^a$, $c^a$, and $\bar c^a$ the Nakanishi-Lautrup, ghost, and antighost fields, respectively.
The gauge-fixed Lagrangian \eqref{eq:lagFP} possesses a formal invariance under gauge transformations of the dynamical fields together with the background field (which thus relate theories in different gauges). The same is true for the corresponding effective action: $\Gamma_{\bar A}[A]=\Gamma_{\bar A^U}[A^U]$, with $A^U$ the gauge transformation of $A$ under $U$. It is convenient to introduce the {\em background field effective action} $\tilde \Gamma[\bar A]=\Gamma_{\bar A}[\bar A]$: First, it is a gauge-invariant functional of its argument and, second, its extrema $\bar A_{\rm ext}$ correspond to actual physical states, that is, to extrema of an actual effective action, namely, those of $\Gamma_{\bar A_{\rm ext}}[A]$~\cite{Abbott:1980hw,Reinosa:2014ooa}. 

In principle, we wish to extremize the background field effective action in the presence of $q\bar q$ sources. Aiming at a simple analytic description, we exploit the fact that because the energy of a flux tube configuration grows linearly with the $q\bar q$ separation, it eventually comes to dominate the total energy. We thus focus on the region of space between (and far enough) the quark and the antiquark, where we can exploit the cylindrical symmetry of the problem. We account for the presence of the sources by imposing a fixed value for the chromoelectric flux trough an infinite plane perpendicular to the $q\bar q$ axis, $\phi^a=\int d^2x_\perp \left<E^a_z\right>$.

Working with an arbitrary background is, of course, too difficult and we shall restrict to a simple subspace designed to capture the desired physics. We introduce $x=(\tau,z,r,\varphi)$, with $\tau$ the Euclidean time, $z$ the coordinate along the $q\bar q$ axis, and $(r,\varphi)$ the polar coordinates in the transverse plane. Lattice simulations show that the chromomagnetic fields are essentially zero whereas the  chromoelectric field has a dominant longitudinal component and a somewhat suppressed transverse component~\cite{Baker:2018mhw}. With this in mind, we choose the ansatz
\begin{equation}\label{eq:bkd}
 \bar A_\mu^a(x)=\frac{\kappa}{g}\delta_{\mu,z}\delta^{a,3}+iA(r)\delta_{\mu,\tau}\delta^{a,1},
\end{equation}
which consists of a (nonperturbative) longitudinal  homogeneous contribution $\kappa/g$ with a fixed color orientation\footnote{Not surprisingly, a nontrivial Wilson loop requires large field configurations $g\bar A\sim 1$. This is analogous to the large background field that encodes the nontrivial Polyakov loop and the associated physics of the confinement-deconfinement transition at nonzero temperature \cite{Weiss:1980rj,Braun:2007bx,Reinosa:2014ooa}.} and a (perturbative) temporal inhomogeneous contribution $A(r)$ in a perpendicular color direction. 
Both $\kappa$ and $A(r)$ are real so that the $z$ component of the average field is real and the $\tau$ component is imaginary, as required in the Euclidean formulation. The nonvanishing components of the background field strength tensor $\bar F_{\mu\nu}^a$ are, in terms of the (real) Minkowski chromoelectric field, $\bar E_z^2(r)=i\bar F^2_{\tau z}=\kappa A(r)$ and $\bar E_r^1(r)=i\bar F^1_{\tau r}=A'(r)$. At the background level, a purely non-Abelian longitudinal chromoelectric field exists for $\kappa\neq0$ and a nontrivial flux tube profile necessarily comes with a nonvanishing radial component.

To make further progress, we expand the background field effective action in powers of the inhomogeneous component $A$, treating the $\kappa$ dependence exactly. We have, up to an irrelevant constant\footnote{The background field gauge symmetry mentioned above guarantees that the $A$-independent term in  $\tilde\Gamma[\bar A]$ cannot depend on the homogeneous background $\kappa$, which is a pure gauge configuration.}, 
\begin{equation}\label{eq:fieldexp}
 \tilde\Gamma[\bar A]=-\frac{1}{2}\int_{x,y}\tilde\Gamma^{(2)}_\kappa(x-y)A(r_x)A(r_y)+{\cal O}(A^4),
\end{equation}
where $\int_x=\int d^4x$. The negative sign in front of the second term on the right-hand side comes from the factor $i$ in the definition \eqref{eq:bkd} and relates to the fact that for imaginary fields, one has to maximize rather than minimize the (Euclidean) effective action. In other words, we have to maximize in $A$ but minimize in $\kappa$. 
The two-point function $\tilde\Gamma^{(2)}_\kappa(x-y)$ in the presence of the homogeneous background field $\kappa$ is invariant under spacetime translations and can be written in terms of its Fourier transform as
\begin{equation}
 \tilde\Gamma^{(2)}_\kappa(x)=\int\frac{d^4q}{(2\pi)^4} e^{-iq\cdot x}\hat\Gamma^{(2)}_\kappa(q)=\hat\Gamma^{(2)}_\kappa(i\partial_x)\delta^{(4)}(x).
\end{equation}
Factorizing a volume $\int_0^Td\tau\int_0^R dz\int_0^{2\pi}d\varphi$, we write $\tilde\Gamma[\bar A]=2\pi RT\,\tilde\gamma_{\kappa}[A]$, with the effective one-dimensional functional\footnote{With a slight abuse of notation, we write $\hat\Gamma^{(2)}_\kappa(0,0,i\partial_r,0)\equiv\hat\Gamma^{(2)}_\kappa(i\partial_r)$.}
\begin{equation}\label{eq:effaction}
 \tilde\gamma_{\kappa}[A]=-\frac{1}{2}\int_0^\infty rdr A(r)\hat\Gamma^{(2)}_\kappa(i\partial_r)A(r)+{\cal O}(A^4).
\end{equation}
The extremization with respect to $A$ yields
\begin{equation}\label{eq:eom}
 \hat\Gamma^{(2)}_\kappa(i\partial_r)A(r)=0
\end{equation}
and the flux constraint reads, at leading order in a semiclassical expansion around the background,
\begin{equation}\label{eq:flux}
 \int_0^\infty rdr A(r)=\frac{\phi}{2\pi \kappa}.
\end{equation}
Flux tube boundary conditions require both $\bar E_z^2$ and $\bar E_r^1$ to be regular at $r=0$ and cylindrical symmetry further implies $\bar E^1_r(0)=0$. We thus impose the regularity conditions $A(0)={\rm const}$ and $A'(0)=0$. The constraint \eqref{eq:flux} implies that $A(r)$ decreases fast enough at $r\to\infty$.

\section{flux tube solution}
We now come to the calculation of $\hat\Gamma^{(2)}_\kappa(q)$. Background field gauge invariance ensures that the nonzero background $\kappa$ simply amounts to a shift $q_z\to q_z \pm \kappa$ for the color component relevant here. In fact, we have $\hat\Gamma^{(2)}_\kappa(q)=\sum_{\sigma=\pm}\hat\Gamma^{(2)}_{\kappa=0}(q_\tau,q_z+\sigma \kappa,\vec q_\perp)/2$, where $\hat\Gamma^{(2)}_{\kappa=0}(q)=\hat\Gamma^{(2)}(q^2)$ only depends on $q^2$. For the case of interest here, $q_\tau=q_z=q_\varphi=0$, we have, in Eq.~\eqref{eq:eom}, $\hat\Gamma^{(2)}_\kappa(i\partial_r)=\hat\Gamma^{(2)}(-\Delta_r+\kappa^2)$, with $\Delta_r=\frac{d^2}{dr^2}+\frac{1}{r}\frac{d}{dr}$.
The tree-level expression $\hat\Gamma^{(2)}(q^2)=q^2$ yields the equation 
\begin{equation}
A''+\frac{A'}{r}-\kappa^2A=0.
\end{equation} 
The solution that vanishes at $r\to \infty$ and satisfies the constraint \eqref{eq:flux} is 
\begin{equation}\label{eq:soltree}
A(r)=\frac{\phi\kappa}{2\pi}K_0(\kappa r),
\end{equation} 
with $K_n(x)$ the modified Bessel function of the second kind of order $n$. The latter is, however, singular at $r\to0$, which results in an infinite energy (see below) for $\kappa\neq0$.

To obtain a regular solution, we take into account the one-loop corrections to the effective action \eqref{eq:fieldexp}. A straightforward calculation with the Lagrangian \eqref{eq:lagFP} gives (see Appendix~\ref{app:loop})
\begin{equation}\label{eq:gammaFP}
 \hat\Gamma^{(2)}_{\rm FP}(q^2)=q^2\left(1+\frac{11}{3}\lambda\ln\frac{q^2}{\mu^2}+c_0\lambda\right),
\end{equation}
where $\lambda=g^2N/(16\pi^2)$, $\mu$ is an arbitrary renormalization scale, and where the number $c_0$ depends on the renormalization scheme. In addition to the trivial root $q^2=0$ (a consequence of the background field gauge invariance), the term in parentheses gives a second root. This is very much welcome since, as discussed below, it is then possible to obtain a solution regular enough at $r=0$ to produce a finite string tension. The caveat here is that the second root is near the Landau pole of the FP theory, where the perturbative expression \eqref{eq:gammaFP} is not reliable. 

{As explained above, a variety of infrared safe continuum approaches have been proposed that avoid this spurious artifact of standard perturbation theory.} Here, we use one of the simplest such approaches, based on the Curci-Ferrari (CF) model \cite{Curci:1976bt}, a simple infrared deformation of the gauge-fixed Lagrangian \eqref{eq:lagFP},
\begin{equation}
 {\cal L}_{\rm CF}={\cal L}_{\rm FP}+\frac{m^2}{2}a_\mu^aa_\mu^a.
\end{equation}
Such a mass term does not spoil the background field gauge invariance \cite{Reinosa:2014ooa} and leaves the ultraviolet sector of the theory unchanged, while successfully screening the FP Landau pole. The model possesses infrared safe renormalization group trajectories, where the coupling remains moderate, allowing for controlled perturbative calculations all the way down to deep infrared scales in line with lattice results \cite{Tissier:2010ts,Pelaez:2021tpq}. Notice though that the presence of massless degrees of freedom---a generic feature of local formulations---still allows for nontrivial infrared physics \cite{Tissier:2010ts,Barrios:2022hzr}, notably, the existence of a confining phase at low temperatures and of a confinement-deconfinement transition \cite{Reinosa:2014ooa}. Here, we show that this also yields controlled semiclassical flux tube solutions.

The one-loop expression of the background field two-point function is presented in Appendix~\ref{app:loop}. In addition to the root $q^2=0$, it presents a single other root in the Euclidean domain, $q^2=M^2>0$. This is a direct consequence of the nonmonotonicity of the background field two-point function at low momentum, which, in the present calculation originates from the loop of massless ghosts. We expect this to be a generic feature of any reliable infrared safe approach, directly related to the positivity violation measured in the lattice Landau gauge gluon correlator. Of course, the precise value of $M$ depends on the details of the infrared safe approach---for instance, in the present CF model, it depends on the values of the parameters $m$ and $\lambda$ and on the renormalization scheme---but we expect its very existence and thus the scenario we propose here to be generic and largely independent of these details. 

In terms of Eq.~\eqn{eq:eom}, the two roots $q^2=0$ and $q^2=M^2$ correspond to taking linear combinations of solutions of the two differential equations $(-\Delta_r+\kappa^2)A=0$ and $(-\Delta_r+\kappa^2)A=M^2A$, that is,
\begin{align}
 A''+\frac{A'}{r}-\kappa^2A=0 \quad{\rm and}\quad A''+\frac{A'}{r}-\tilde\kappa^2A=0,
\end{align}
where we defined $\tilde \kappa=\sqrt{\kappa^2-M^2}$.  
 There exists a unique\footnote{There exists no solution compatible with the constraint \eqref{eq:flux} in the case $\kappa\le M$. }  linear combination that vanishes at $r\to\infty$ and is finite at $r=0$, namely, after imposing the flux constraint \eqref{eq:flux},
\begin{equation}\label{eq:sol1}
 A(r)=\frac{\phi}{2\pi\kappa}\frac{\kappa^2\tilde\kappa^2}{M^2}\left[K_0(\tilde\kappa r)-K_0(\kappa r)\right].
\end{equation}
This is the central result of the present work: the nonmonotonicity of the background field two-point function, an expected fundamental feature of YM theories, guarantees the existence of a nontrivial flux tube profile. We shall see below that Eq.~\eqref{eq:sol1} describes well the existing lattice data for the latter.

As mentioned above, the precise value of $M$ in Eq.~\eqref{eq:sol1} depends on the details of the infrared safe approach. However, the generic features of our scenario happen to be largely independent of these details and thus robust. To illustrate this in the spirit of the present perturbative calculation, it suffices to consider the low $q^2$ expansion (see Appendix~\ref{app:loop})
\begin{equation}\label{eq:LO}
 \frac{\hat\Gamma^{(2)}(q^2)}{q^2}=1+\frac{\lambda}{6}\left[\ln\frac{q^2}{m^2}+c_m+{\cal O}\!\left( \frac{q^2}{m^2}\ln\frac{q^2}{m^2}\right)\right].
\end{equation}
All the terms in brackets but $c_m=c(m/\mu)$ are independent of the renormalization scheme. This gives a root at the (model-dependent) infrared scale $M^2=m^2\exp(-6/\lambda-c_m)$. In contrast with the FP case \eqref{eq:gammaFP}, where the infrared and the ultraviolet logarithms  $\ln q^2$ and $\ln\mu^2$ are intrinsically entangled, the $\ln (q^2/m^2)$ term in Eq.~\eqref{eq:LO} is of purely infrared origin. The positive sign of the coefficient of that logarithm is the crucial point here, which guarantees the existence of the root $M^2>0$. 

 For small couplings, we typically have $M^2\ll\kappa^2$ and Eq.~\eqref{eq:sol1} approximates to
\begin{equation}\label{eq:sol}
 A(r)\approx\frac{\phi \kappa}{4\pi}f(\kappa r)  \quad{\rm with}\quad f(x)=xK_1(x).
\end{equation}
Remarkably, the solution \eqref{eq:sol} is independent of $M$ and thus of the details of the infrared safe approach. Also, note that it survives the limit $\lambda\to0$, although the one-loop correction to the effective action has been crucial to establish it. In this sense, it is the leading order of an approximation scheme with corrections that can be computed in a systematic expansion around Eqs.~\eqref{eq:fieldexp} and \eqref{eq:LO}. For the purpose of illustrating our scenario in the simplest way, we shall use the expression \eqref{eq:sol} in our discussion below. We have also performed calculations with the complete expression \eqref{eq:sol1} and the actual value of $M^2$ obtained from the complete one-loop expression of $\hat\Gamma^{(2)}(q^2)$ given in Appendix~\ref{app:loop}. We mention how the results are affected when needed.

The energy of the field configuration \eqref{eq:sol} is 
\begin{equation}
 V_{q\bar q}(R)=\frac{1}{2}\int d^3x\left<E^2+B^2\right>,
\end{equation}
where the integral spans the whole space.  Assuming that for large $q\bar q$ separation $R$, the total energy is dominated by the region between the sources, we have\footnote{At large (Euclidean) time $T$, the Wilson loop ${\cal W}(T,R)$ describing a $q\bar q$ pair separated by a distance $R$ behaves as ${\cal W}(T,R)\propto\exp(-V_{q\bar q}(R)T)$ \cite{Michael:1986yi}. A linear behavior $V_{q\bar q}(R)\propto R$ is equivalent to the area law for the Wilson loop.} $V_{q\bar q}(R) \approx\sigma R$ where the energy per unit length $\sigma$ defines the string tension. At leading order, we obtain
\begin{align}\label{eq:sigma}
 \sigma= \pi \int_0^\infty r dr\left[(A')^2+\kappa^2A^2\right] \approx\frac{\phi^2\kappa^2}{16\pi}.
\end{align}
Note that this expression is independent of the details of the infrared safe approach and thus generic.\footnote{\label{foot}The complete expression \eqref{eq:sol1}, yields $$\sigma=\frac{\phi^2\kappa^2}{8\pi}\left(\frac{\kappa^2}{M^2}-1\right)^2\left[\ln\left(1-\frac{M^2}{\kappa^2}\right)+\frac{M^2}{\kappa^2-M^2}\right].$$}

Before proceeding with the discussion, let us mention a caveat of our flux tube solution \eqref{eq:sol1} or \eqref{eq:sol}. The key point of the present scenario is that the loop contribution from the massless degrees of freedom to the background field two-point function in Eq.~\eqref{eq:fieldexp} smoothens the singularity of the tree-level solution \eqref{eq:soltree} at $r\to0$. However, we point out that this mechanism alone is not enough to produce a completely regular solution for we have $A''(r)\propto\ln (\kappa r)$ for $r\to0$. Such a singularity is, of course, spurious as there is nothing special about the line $r=0$ and a fully consistent flux tube solution must be completely regular there. 

This difficulty is to be expected for linearized equations of motion and is thus a mere artifact of our quadratic approximation \eqref{eq:fieldexp} for the effective action. Physically, we expect the latter to be valid far from the core of the flux tube where the field is small but nonlinearities cannot be neglected well inside the flux tube. It is worth emphasizing though that the present approximation is sufficient to give a reasonable estimate of the shape of the flux tube and of the corresponding string tension as we discuss below. This indicates that the range of values of $r$ where nonlinearities are important is small. We proceed with this analysis and leave the discussion of the regularity near $r\to 0$ and the role of the higher-than-quadratic terms in the action \eqref{eq:fieldexp} for a separate work.

\begin{figure}[t!]
  \centering
  \includegraphics[width=.95\linewidth]{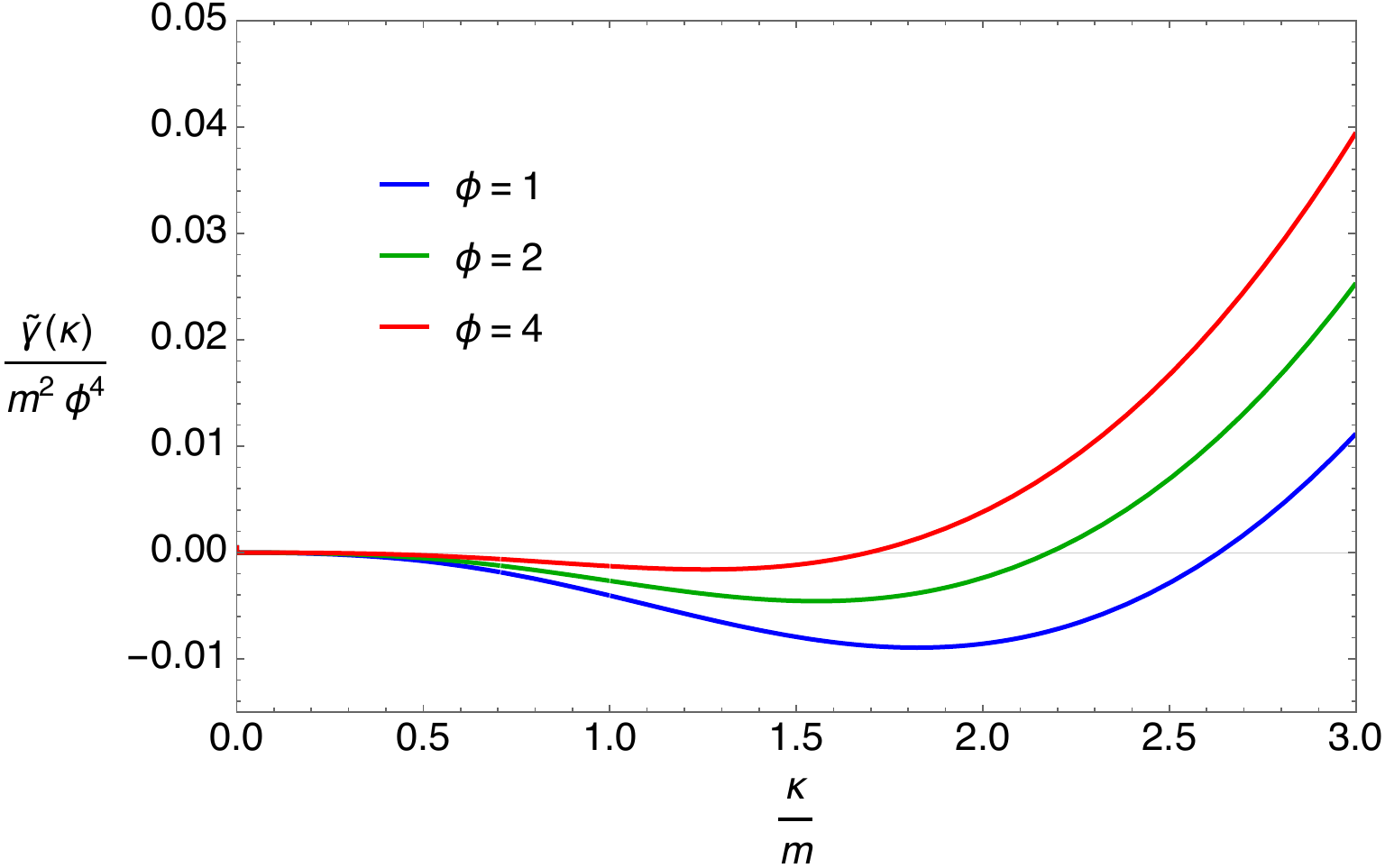}
  \caption{The effective potential \eqref{eq:gammaK} (arbitrary normalization) for the background $\kappa$ for various values of the flux $\phi$ for $N=2$.}
  \label{fig:gammatilde}
\end{figure}

\section{The homogeneous background}
The homogeneous background field component $\kappa$ is dynamically determined as the minimum of the function $\tilde\gamma(\kappa)$, defined as the functional \eqref{eq:effaction} evaluated at the extremum \eqref{eq:sol1} or \eqref{eq:sol}. However, our quadratic approximation trivially yields $\tilde\gamma(\kappa)=0$ and there is no preferred value of $\kappa$. We lift this spurious degeneracy by including the next terms in the field expansion \eqref{eq:effaction}. To simplify matters, we treat the higher-than-quadratic terms in a local potential approximation, that is,
\begin{equation}\label{eq:LPA}
 \tilde\gamma_{\kappa}[A]=\int_0^\infty\! rdr \!\left[-\frac{1}{2}A\hat\Gamma^{(2)}_\kappa(i\partial_r)A+\tilde V_\kappa(A)\right]\!.
\end{equation}
Using the equation $\hat\Gamma^{(2)}_\kappa(i\partial_r)A_\kappa=\tilde V_\kappa'(A_\kappa)$, defining the extremum $A_\kappa(r)$, we obtain, for $\tilde\gamma(\kappa)=\tilde\gamma_\kappa[A_\kappa]$,
\begin{equation}\label{eq:gammamin}
 \tilde\gamma(\kappa)=\int_0^\infty\! rdr \left[-\frac{1}{2}A_\kappa\tilde V'_\kappa(A_\kappa)+\tilde V_\kappa(A_\kappa)\right].
\end{equation}
This expression being of the order of the degeneracy-lift potential $\tilde V_\kappa$, one can use expression \eqref{eq:sol1} or \eqref{eq:sol} for $A_\kappa$ to compute the minimum $\kappa_0$ at leading order. 

\begin{figure}[t!]
  \centering
  \includegraphics[width=.85\linewidth]{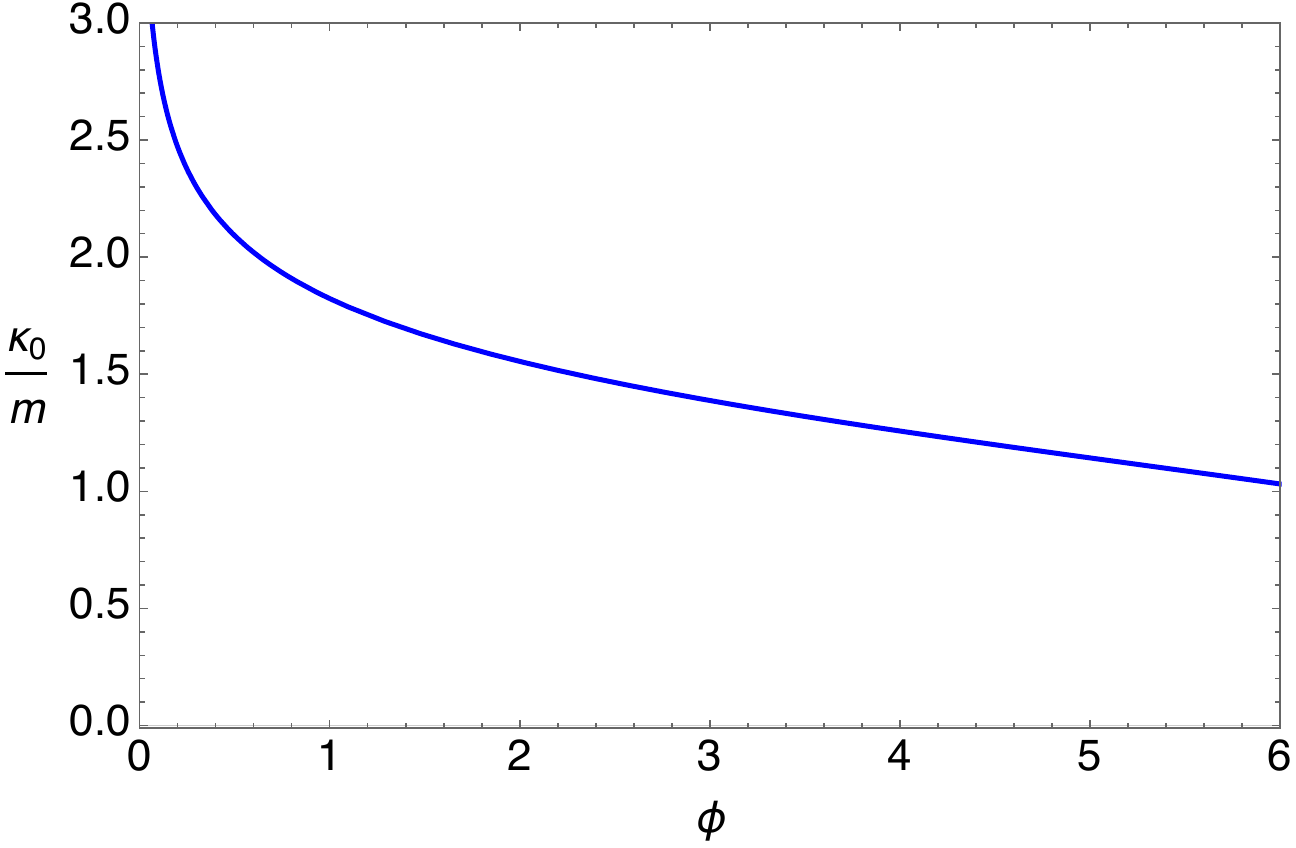}
  \caption{The parameter $\kappa_0/m$ as a function of $\phi$ for $N=2$.}
  \label{fig:Kmin}
\end{figure}

As before, a controlled infrared safe approach is required to compute the potential $\tilde V_\kappa(A)$ which, due to infrared effects, is nonanalytic in $A$. Its field expansion reads, at one-loop order in the CF model,
 \begin{align}\label{eq:A4}
\tilde V_\kappa(A)=a_{\kappa}A^4\left(\ln \frac{A}{m}+b_\kappa\right)+{\cal O}(A^6\ln A),
 \end{align}
where the functions $a_\kappa=a(\kappa/m)$ and $b_\kappa=b(\kappa/m)$ are given in Appendix~\ref{app:A4} for $N=2$ and $N=3$. Using the profile \eqref{eq:sol}, Eqs.~\eqref{eq:gammamin}, and \eqref{eq:A4} yield
\begin{equation}\label{eq:gammaK}
 \tilde\gamma(\kappa)=-\frac{\phi^4\kappa^2}{(4\pi)^4}a_\kappa\left[c_1 \left(\ln \frac{\phi \kappa}{4\pi m}+\frac{1}{2}+b_\kappa\right)+c_2\right],
\end{equation}
with the constants $c_1=\int_0^\infty dx xf^4(x)\approx0.24$ and $c_2=\int_0^\infty dx xf^4(x)\ln f(x)\approx-0.08$. In Fig.~\ref{fig:gammatilde}, we show the function $\tilde\gamma(\kappa)$ for various values of $\phi$, which exhibits a nontrivial minimum\footnote{We find that $\kappa_0(\phi)$ presents a discontinuous jump to $\kappa_0=0$ for large values of $\phi$ for both $N=2$ and $N=3$. However, such a limiting value of the flux appears to be an artifact of the approximation \eqref{eq:sol}, whose validity becomes questionable at small $\kappa$, corresponding to large $\phi$. We find a nontrivial minimum $\kappa_0\neq0$ for all values of $\phi$ when using the expression \eqref{eq:sol1} for the flux tube profile.} $\kappa_0\neq0$ that we plot as a function of $\phi$ in Fig.~\ref{fig:Kmin}.

\section{Discussion}
At this point, we are in a position to predict the string tension $\sigma$ in terms solely of the flux $\phi$. It is interesting to attempt a rough order-of-magnitude estimate from a simplistic perturbative picture, where the flux is directly given by the color charge of a static quark, that is,  $\phi=g\sqrt{(N-1)/(2N)}$ in the fundamental representation of SU($N$). For our estimate, we use the values of the coupling and the mass obtained by fitting the Landau gauge gluon and ghost propagators measured in numerical simulations against the one-loop CF predictions \cite{Tissier:2010ts}. For the SU(2) data of Ref.~\cite{Cucchieri:2007rg}, the best fit values are $g=7.5$ ($\lambda=0.71$) and $m=0.68~{\rm GeV}$ at the scale $\mu=1~{\rm GeV}$. This gives $\phi=g/2=3.75$, which corresponds to $\kappa_0/m=1.3$, that is, $\kappa_0=0.87~{\rm GeV}$, roughly consistent with the scale $\mu$. We get $\sqrt\sigma=463~{\rm MeV}$, to be compared with  $\sqrt{\sigma_{\rm lat}}=440~{\rm MeV}$ used in Ref.~\cite{Cucchieri:2007rg}. Similar fits of the SU(3) data of Ref.~\cite{Dudal:2010tf} give $g=4.9$ ($\lambda=0.46$) and $m=0.54~{\rm GeV}$. We thus have $\phi=g/\sqrt3=2.83$, for which we find $\kappa_0/m=1.55$, that is, $\kappa_0=0.84~{\rm GeV}$, resulting in $\sqrt\sigma=334~{\rm MeV}$, to be compared, again, with $\sqrt{\sigma_{\rm lat}}=440~{\rm MeV}$ \cite{Dudal:2010tf}. 

Repeating this analysis with the complete expression \eqref{eq:sol1} for the flux tube profile gives similar results. Using the renormalization scheme of Ref.~\cite{Tissier:2010ts}, for which the best fit values quoted above are obtained, we find, for the SU(2) case, $M=0.7~{\rm GeV}$ (see Appendix~\ref{app:loop}). With this value and using $\phi=g/2=3.75$, the potential \eqref{eq:gammamin} computed with the profile \eqref{eq:sol1} is minimized for $\kappa_0/m=1.56$, that is, $\kappa_0=1.06~{\rm GeV}$. The relevant expression of the string tension (see footnote~\ref{foot}) then gives $\sqrt\sigma=459~{\rm MeV}$. For the SU(3) case, we obtain, with the parameters quoted above, $M=0.58$, $\kappa_0/m=1.8$, that is, $\kappa_0=0.972~{\rm GeV}$, and $\sqrt\sigma=333~{\rm MeV}$. 

These relatively good numbers must be taken with a grain of salt, first, because of the naive estimation of the flux and, second, because of the expected precision of the present leading-order calculation. We have repeated the analysis using different renormalization schemes for the definition of the CF parameters $m$ and $\lambda$, which roughly yields a 30\% variability in the above estimations.\footnote{For instance, in the so-called infrared safe scheme, we find the best fit parameters for SU(2): $g=6.2$ ($\lambda=0.49$) and $m=0.5~{\rm GeV}$ at the scale $\mu=1~{\rm GeV}$ and we obtain $M=0.72~{\rm GeV}$. The approximation \eqref{eq:sol} gives $\kappa_0=0.69~{\rm GeV}$ and $\sqrt\sigma=300~{\rm MeV}$ whereas the complete expression \eqref{eq:sol1} yields $\kappa_0=0.93~{\rm GeV}$ and $\sqrt\sigma=293~{\rm MeV}$.}

\begin{figure}[t!]
  \centering
  \includegraphics[width=.9\linewidth]{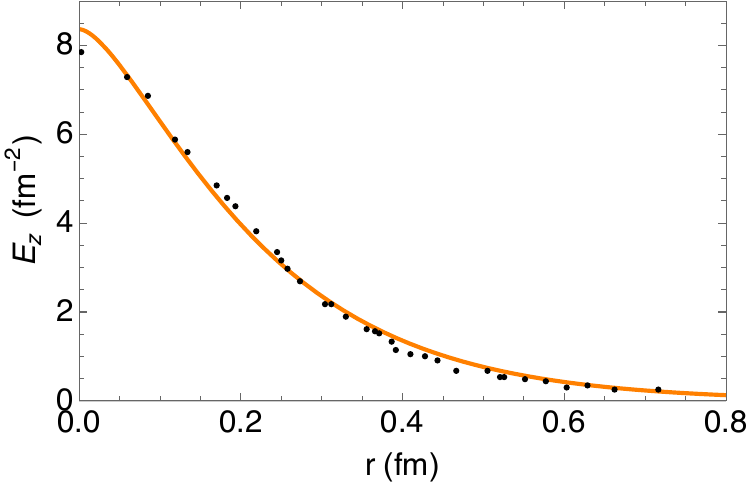}
  \caption{The lattice data of Ref.~\cite{Koma:2003hv} for the longitudinal chromoelectric field against the present leading-order result $\bar E^{a=2}_z(r)=\kappa A(r)=\phi \kappa^2 f(\kappa r)/(4\pi)$.}
  \label{fig:Ez}
\end{figure}

These qualitative estimates are supported by actual fits of the flux tube profile obtained in lattice simulations as we now discuss.
Figure~\ref{fig:Ez} shows a fit of the SU(2) data of Ref.~\cite{Koma:2003hv} for the longitudinal chromoelectric field against the present leading-order expression\footnote{We obtain equally good fits for the more recent data of Refs.~\cite{Cea:2012qw,Cea:2014uja} for both SU(2) and SU(3). However the unconventional normalization of the operator used there as a proxy for the field does not allow for a direct estimate of the string tension \cite{Cea:2012qw,Battelli:2019lkz}.}. The best fit values using the approximation\footnote{We have checked that either the complete expression \eqref{eq:sol1} or the approximation \eqref{eq:sol} give fits of equally good quality with similar values of $\phi$ and $\kappa$ and, thus, of $\sigma$. For instance, for the SU(2) case with $M=0.7~{\rm GeV}$, we obtain $\phi=2.39$ and $\kappa=1.42~{\rm GeV}$, yielding $\sqrt\sigma=437~{\rm MeV}$.} \eqref{eq:sol} are  $\phi=2.38$ and $\kappa=1.33~{\rm GeV}$. This yields $\sqrt\sigma=446~{\rm MeV}$, remarkably close to the value $\sqrt{\sigma_{\rm lat}}=440~{\rm MeV}$ used in Ref.~\cite{Koma:2003hv}. 
For this value of $\phi$, we obtain $\kappa_0/m=1.5$, which, when combined with the value of $\kappa$ from the fit, gives a CF mass $m= 0.9~{\rm GeV}$, in qualitative agreement with estimates from other sources~\cite{Pelaez:2021tpq}. 

It is interesting to make predictions independent of the details of the infrared safe approach, here, the CF mass. For instance, the critical temperature $T_c$ of the confinement-deconfinement transition is well measured on the lattice~\cite{Lucini:2012gg}. The perturbative CF model yields, at leading order, $T_c/m=0.34$ for $N=2$~\cite{Reinosa:2014ooa}, from which we obtain $T_c/\sqrt\sigma=(T_c/m)(m/\kappa)(\kappa/\sqrt\sigma)= 0.68$, in remarkable agreement with the lattice result $(T_c/\sqrt\sigma)_{\rm lat}=0.7$~\cite{Lucini:2012gg}.

Fits of the SU(3) flux tube profile of Ref.~\cite{DIK:2003alb} are of as good quality as above, with $\phi=2.8$ and $\kappa=1.08~{\rm GeV}$. This gives $\sqrt\sigma=425~{\rm MeV}$, where Ref.~\cite{DIK:2003alb} measures $\sqrt{\sigma_{\rm lat}}=464~{\rm MeV}$. We obtain $\kappa_0/m=1.55$, giving $m=0.7~{\rm GeV}$ which, again, agrees qualitatively with other estimates~\cite{Pelaez:2021tpq}. The leading-order CF prediction for the transition temperature is  $T_c/m=0.36$ \cite{Reinosa:2014ooa} and we obtain $T_c/\sqrt\sigma=0.59$, where $(T_c/\sqrt\sigma)_{\rm lat}=0.64$~\cite{Lucini:2012gg}.

\section{Conclusion}
In conclusion, we have proposed a generic scenario for explaining the existence of a flux tube configuration solely in terms of basic infrared properties of YM dynamics in the Landau gauge, namely, the moderate infrared coupling, the nonzero gluon screening mass and the associated positivity violations reflecting the remaining massless modes. This is a nontrivial extension of the observation that these infrared aspects alone suffice to explain the confinement of static quarks---as measured by a vanishing Polyakov loop---at low temperatures~\cite{Braun:2007bx}. We have illustrated our point with the perturbative CF model, which, among other possible infrared safe approaches, allows for a simple and transparent analytical treatment. We have obtained consistent semiclassical flux tube solutions that describe well the existing lattice data and correctly predict the string tension for a static $q\bar q$ pair. Although obtained within a definite infrared safe approach, we have argued our results to be robust and to describe a fundamental aspect of the physics of confinement. It is of great interest to test the present scenario in other infrared safe approaches.

We have mentioned that the mechanism proposed here is not enough to produce a completely regular solution at $r\to0$ and it is important to resolve this issue. As we have argued, this requires one to go beyond the quadratic approximation \eqref{eq:fieldexp} and the corresponding simple analytical approximation \eqref{eq:sol1}. A first study in this direction consists in solving the nonlinear equations of motion for the flux tube profile stemming from the local potential approximation \eqref{eq:LPA} with the quartic potential \eqref{eq:A4}. A preliminary study indicates that this solves the regularity issue and provides fully regular profiles. Remarkably, the nonanalytic contribution $\propto A^4\ln (A/m)$ from the massless degrees of freedom of the infrared safe approach appears essential in producing regular flux tube solutions, thereby extending the scenario proposed in this article. We will present this analysis elsewhere.  

The present work opens many interesting directions of research, among which, studying the string tension for sources in higher representations \cite{Bali:2000un}, computing the corrections to the present semiclassical result, studying the finite $q\bar q$ distance effects and the L\"uscher term \cite{Luscher:1980ac}, including dynamical quarks \cite{DIK:2003alb} as well as nonzero temperature effects \cite{DiGiacomo:1990hc,Cea:2015wjd}.

\section*{Acknowledgements}
We thank B. Delamotte, D. Dudal, and G. Hern\'andez for useful discussions.
M. P. and N. W. are thankful for the support of the Programa de  Desarrollo de las Ciencias B\'asicas (PEDECIBA). This work
received support from the French-Uruguayan Institute
of Physics Project (IFUP) and from the Agencia Nacional de Investigaci\'on e Innovaci\'on (Uruguay), Grant No
FCE-1-2021-1-166479.

\vspace{2cm}

\appendix

\section{Two-point function}\label{app:loop}

The background field two-point function $\hat\Gamma^{(2)}(q^2)$ can be computed at one-loop order in the CF model by straightforward diagrammatic techniques. Equivalently, it can be obtained from the field expansion of the background field effective action $\tilde\Gamma[\bar A]$ evaluated for homogeneous fields $\kappa$ and $A$. The coefficient of the $A^2/2$ term is $\hat\Gamma^{(2)}_{\kappa}(q=0)=\hat\Gamma^{(2)}(\kappa^2)$, as explained in the text. We have performed both calculations as a cross-check. The result is ultraviolet divergent and requires renormalization. This is achieved in terms of the background field renormalization factor $Z_{\bar A}$, which relates the bare and renormalized background fields as $\bar A=\sqrt{Z_{\bar A}}\bar A_R$. The gauge invariance of the background field action guarantees that the product $g\bar A$ is finite. We can thus choose the renormalization condition $Z_{g^2}Z_{\bar A}=1$. Using dimensional regularization with $d=4-2\epsilon$ and defining the dimensionless renormalized coupling $g_R$ through $g^2=Z_{g^2}g^2_R\mu^{2\epsilon}$, with $\mu$ the renormalization scale, we have, at one-loop order, 
\begin{equation}
Z_{g^2}=1-\frac{11\lambda}{3\epsilon}+\lambda\delta z_f,
\end{equation}
where $\lambda=g_R^2N/(16\pi^2)$. The finite part $\delta z_f$ is a function of the dimensionless ratio $m/\mu$ with a regular limit $m/\mu\to0$. We obtain, writing $\bar\mu^2=4\pi\mu^2 e^{-\gamma}$, with $\gamma$ the Euler-Mascheroni constant, 
\begin{equation}\label{appeq:g2}
 \hat\Gamma^{(2)}(q^2)=q^2\left(1-\lambda\delta z_f+\frac{11}{3}\lambda\ln\frac{m^2}{\bar\mu^2}\right)+\lambda m^2{\cal F}\left(\frac{q^2}{m^2}\right),
\end{equation}
where 
\begin{align}
{\cal F}(x)&=4-\frac{223}{36}x\nonumber\\
\label{suppeq:FF}
 &-\frac{x}{12}\left(x^2-20x+12\right)\left(\frac{x+4}{x}\right)^{3/2}\tanh^{-1}\sqrt{\frac{x}{x+4}}\nonumber\\
 &+\frac{x+1}{12} \left(x^2-10x+1\right) \ln\left(1 + x\right)\nonumber\\
 &-\frac{x}{24}\left(x^2-4x-4\right) \ln x.
\end{align}
The corresponding expression \eqref{eq:gammaFP} for the FP theory is obtained by taking the limit $m\to 0$ and using the large-$x$ expansion
\begin{equation}
 {\cal F}(x)=\frac{11}{3}x\ln x-{205\over36}x+{\cal O}(\ln x) .
\end{equation}

One checks that $\hat\Gamma^{(2)}(q^2=0)=0$, as required by the background gauge invariance. Moreover, thanks to the last term $\propto \ln x$ in Eq.~\eqref{suppeq:FF}, which originates from the remaining massless degrees of freedom of the (infrared regulated) theory, there always exists---whatever the choice of renormalization condition---a unique root $\hat\Gamma^{(2)}(q^2)=0$ at $q^2\neq0$. As long as the coefficient of the term in the parentheses $\propto q^2$ on the right-hand side of Eq.~\eqref{suppeq:FF} is positive, this root lies at $x<1$. In fact, in the spirit of the perturbative expansion, where the mentioned $q^2$-term is $1+{\cal O}(\lambda)$, this root lies at $x\ll1$. In this limit, we have
\begin{equation}
 {\cal F}(x)=\frac{1}{6}x\ln x-{7\over9}x+{\cal O}(x^2\ln x).
\end{equation}

\begin{figure}[t!]
  \centering
  \includegraphics[width=.9\linewidth]{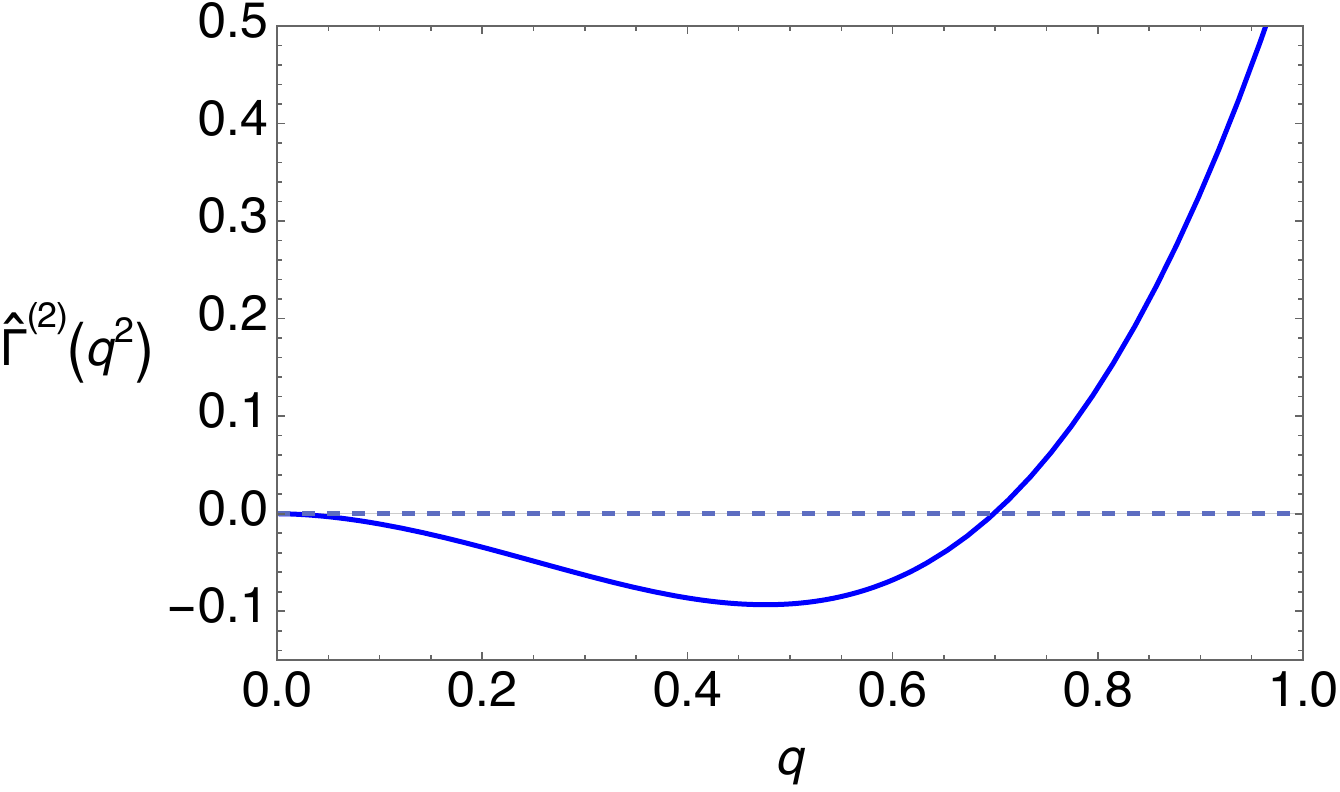}
  \caption{The function \eqref{appeq:g2} with the counterterm \eqref{appeq:ct} for the SU(2) parameters $m=0.68~{\rm GeV}$ and $g=7.5$ at $\mu=1~{\rm GeV}$. The nonmonotonous behavior at low momentum is related to the positivity violation of the gluon propagator in the Landau gauge---corresponding to a vanishing background---and is responsible for the nontrivial root at $q=M=0.7~{\rm GeV}$.}
  \label{fig:Gammatilde2}
\end{figure}

For definiteness, we give the expression of the counterterm $\delta z_f$ in the so-called vanishing momentum scheme, used in Ref.~\cite{Tissier:2010ts} to fit the lattice data for the ghost and the gluon correlators,
\begin{align}\label{appeq:ct}
 \delta z_f&= \frac{11}{3}\ln\frac{\mu^2}{\bar\mu^2}- \frac{211}{36}+\frac{33}{8y}-\frac{1}{12y^2} \nonumber \\
 &  -\frac{y^2-20y+12}{12} \left(\frac{y+4}{y}\right)^{3/2}\tanh^{-1}\sqrt{\frac{y}{y+4}}\nonumber \\
 & +\frac{y^2-4y+1}{12}  \left(\frac{y+1}{y}\right)^3\ln\left(1+ y\right)\nonumber\\
 & - \frac{y^2+12y + 86}{24} \ln y,
\end{align}
where $y=\mu^2/m^2$.
We plot the corresponding background field two-point vertex function \eqref{appeq:g2} in Fig.~\ref{fig:Gammatilde2} for the case $N=2$. 

\section{Quartic potential}\label{app:A4}

Next, we obtain the potential \eqref{eq:A4} by evaluating the background field effective action for homogeneous fields $\kappa$ and $A$ at one-loop order and expanding in powers of $A$. Because of the massless degrees of freedom, this expansion is nonanalytic and there appears a $A^4\ln A$ term. The calculation is straightforward but cumbersome as it involves two homogenous backgrounds in different spacetime and color directions. We shall present it in detail elsewhere and we simply give the resulting expressions for $N=2$ and $N=3$ here. 

Writing the coefficients of Eq.~\eqref{eq:A4} as 
\begin{align}
 a_\kappa&=\frac{g^4}{128\pi^2}\tilde a_N(\kappa^2/m^2)\\
 a_\kappa b_\kappa&=\frac{g^4}{960\pi^2}\tilde c_N(\kappa^2/m^2),
\end{align}
we have, for $N=2$,
\begin{equation}
 \tilde a_2(x) = 6\left[\frac{1}{ (1 + x)^2}-2\right]
\end{equation}
and
\begin{align}
 \tilde c_2(x)&=\frac{p_1(x)}{16x(1+x)^2}+\frac{p_2(x)}{2(1-x^2)^2}\nonumber\\
 &+\frac{p_3(x)}{8(1+x-2x^2)^2}\ln x\nonumber\\
 &+\frac{p_4(x)}{4x(1-5x^2+4x^4)^2}\ln(1+x)\nonumber\\
 &+\frac{p_5(x)}{64(1-4x^2)^2}\ln(1+4x)\nonumber\\
 &+\frac{p_6(x)}{x^2(1+x)^2}\sqrt{\frac{1+x}{x}}\tanh^{-1}\sqrt{\frac{x}{1+x}}\nonumber\\
 &+\frac{p_7(x)}{4x^{2}(1-2x)^2}\sqrt{\frac{4+x}{x}}\tanh^{-1}\sqrt{\frac{x}{4+x}},
\end{align}
where
\begin{align}
 p_1(x)&=1728+3276x+4144x^2+4567x^3+3190x^4\nonumber\\
 &+1035x^5-112x^6+12x^7\\
 p_2(x)&=(-31 - 62 x + 71 x^2 + 184 x^3 - 108 x^4 - 60 x^5 \nonumber\\
 &+ 70 x^6)\ln2\\
 p_3(x)&=180 + 800 x - 280 x^2 - 2680 x^3 + 1599 x^4 \nonumber\\
 &+ 1987 x^5 - 
 2365 x^6 + 1093 x^7 + 54 x^8 - 112 x^9 \nonumber\\
 &+ 12 x^{10}\\
 p_4(x)&=-4 - 250 x + 541 x^2 + 3425 x^3 - 3222 x^4 \nonumber\\
 &- 15799 x^5 + 
 6189 x^6 + 39091 x^7 + 1628 x^8 \nonumber\\
 &- 51877 x^9 - 68 x^{10} +19754 x^{11} - 1180 x^{12} \nonumber\\
 &+ 268 x^{13} + 400 x^{14} - 48 x^{15}\\
 p_5(x)&=-5x (1 + 4 x)^3 (103 - 168 x + 112 x^2)\\
 p_6(x)&=144 + 232 x - 35 x^2 - 310 x^3 - 177 x^4 + 30 x^5 \nonumber\\
 &+ 35 x^6\\
 p_7(x)&=-576 + 1952 x - 432 x^2 - 3112 x^3 + 452x^4\nonumber\\
 &+3948x^5-2083x^6-1917x^7+362x^8-136x^9\nonumber\\
 &+12x^{10}.
\end{align}

The corresponding expressions for $N=3$ can be written $\tilde a_3(x) =\tilde a_2(x) +\delta \tilde a(x)$ and $\tilde c_3(x) =\tilde c_2(x) +\delta \tilde c(x)$, with
\begin{equation}
 \delta \tilde a(x) =\frac{1}{ (1 + x)^2}-2 
\end{equation}
and
\begin{align}
 \delta\tilde c(x)&=\frac{\delta p_1(x)}{16x(1+x)^2}+\frac{\delta p_2(x)}{2(1+x)^2}+\frac{\delta p_3(x)}{64}\ln x\nonumber\\
 &+\frac{\delta p_4(x)}{32x(1+x)^2}\ln(1+x)\nonumber\\
 &+\frac{\delta p_7(x)}{32x}\sqrt{\frac{4+x}{x}}\tanh^{-1}\sqrt{\frac{x}{4+x}},
\end{align}
where
\begin{align}
 \delta p_1(x)&=720 + 1018 x + 592 x^2 + 971 x^3 + 964 x^4 \nonumber\\
 &+ 286 x^5 - 28 x^6 + 3 x^7\\
 \delta p_2(x)&=15 (1 + 4 x + 2 x^2)\ln2\\
 \delta p_3(x)&=240 + 800 x - 1040 x^2 + 520 x^3 - 9 x^4 \nonumber\\
 &- 50 x^5 + 6 x^6\\
 \delta p_4(x)&=-16 - 792 x - 342 x^2 + 3483 x^3 + 6334 x^4 \nonumber\\
 &+ 2924 x^5 + 58 x^6 + 
 103 x^7 + 38 x^8 - 6 x^9\\ 
 \delta p_7(x)&=-1440 + 880 x + 80 x^2 - 2464 x^3 - 902 x^4 \nonumber\\
 &+ 127 x^5 - 62 x^6 + 6 x^7.
\end{align}

We emphasize that expression \eqref{eq:A4} is ill defined in the FP theory, hence the need for an infrared safe approach. For instance,  for $N=2$, we have, in the limit $m\to0$, $\tilde a_2(x)=-12+{\cal O}(x^{-2})$ and $\tilde c_2(x)=-{2565\over 16}\ln x+{\cal O}(x^{0})$, which gives
\begin{equation}
 \tilde V_\kappa(A)=-\frac{3g^4}{32\pi^2}A^4\left(\ln\frac{A}{m}+\frac{57}{16}\ln\frac{\kappa}{m}\right)+{\cal O}(g^4A^4).
\end{equation}

Finally, in the weak coupling regime, where $\phi\ll1$ and $\kappa_0/m\gg1$, we can approximate the function \eqref{eq:gammaK} as
\begin{equation}
 \tilde\gamma(\kappa)=\frac{3g^4}{32\pi^2}c_1\frac{\phi^4\kappa^2}{(4\pi)^4} \left(\ln \phi+\frac{73}{16}\ln\frac{\kappa}{m}+C\right),
\end{equation}
with $C$ a calculable constant. We get $\kappa_0/m\propto\phi^{-16/73}$.


\end{document}